\documentclass[10pt, conference, compsocconf]{IEEEfocs}
\IEEEoverridecommandlockouts

\ifCLASSINFOpdf
\else
\fi
\newcommand\remove[1]{}

\newcommand{\rnote}[1]{}
\newcommand{\jnote}[1]{}

\newcommand{\E}{\mathcal E}
\newcommand{\lines}{{\rm lines}}

\newcommand{\length}{\operatorname{length}}

\newcommand{\per}{\operatorname{Per}}

\renewcommand{\span}{\operatorname{span}}

\newcommand{\cut}{\operatorname{Cut}}

\newcommand{\1}{\mathbf{1}}
\newcommand{\e}{\varepsilon}
\newcommand{\R}{\mathbb{R}}
\newcommand{\N}{\mathbb{N}}

\newcommand{\Lip}{\mathrm{Lip}}

\renewcommand{\span}{\mathrm{\bf span}}

\renewcommand{\H}{\mathbb H}

\usepackage{amssymb}
\usepackage{amsthm}
\usepackage{amsmath}
\usepackage{hyperref,mathrsfs,txfonts,epsfig,graphicx,graphics}

\newtheorem{theorem}{Theorem}[section]

\newtheorem{remark}{Remark}[section]

\begin{document}
%
\title{A $(\log n)^{\Omega(1)}$ integrality gap for the Sparsest Cut SDP}


\author{\IEEEauthorblockN{Jeff Cheeger\IEEEauthorrefmark{1}\thanks{\IEEEauthorrefmark{1}Research supported in part by NSF grant DMS-0704404.}}
\IEEEauthorblockA{Courant Institute\\
New York University\\
New York, USA\\
Email: cheeger@cims.nyu.edu}
\and
\IEEEauthorblockN{Bruce Kleiner\IEEEauthorrefmark{2}\thanks{\IEEEauthorrefmark{2}Research supported in part by NSF grant DMS-0805939.}}
\IEEEauthorblockA{Courant Institute\\
New York University\\
New York, USA\\
Email: bkleiner@cims.nyu.edu}
\and
\IEEEauthorblockN{Assaf Naor\IEEEauthorrefmark{3}\thanks{\IEEEauthorrefmark{3}Research supported in part by NSF grants DMS-0528387, CCF-0635078 and CCF-0832795, BSF grant 2006009, and the Packard Foundation.}}
\IEEEauthorblockA{Courant Institute\\
New York University\\
New York, USA\\
Email: naor@cims.nyu.edu}
}



%


\maketitle

\begin{abstract}
We show that the Goemans-Linial semidefinite relaxation of the
Sparsest Cut problem with general demands has integrality gap $(\log
n)^{\Omega(1)}$. This is achieved by exhibiting $n$-point metric
spaces of negative type whose $L_1$ distortion is $(\log
n)^{\Omega(1)}$. Our result is based on quantitative bounds on the
rate of degeneration of Lipschitz maps from the Heisenberg group to
$L_1$ when restricted to cosets of the center.
\end{abstract}


%
\IEEEpeerreviewmaketitle

\section{Introduction}
The $L_1$ distortion of a metric space $(X,d)$, commonly denoted
$c_1(X,d)$, is the infimum over $D>0$ for which there exists a
mapping $f:X\to L_1$ such that $\frac{\|f(x)-f(y)\|_1}{d(x,y)}\in
[1,D]$ for all distinct $x,y\in X$. (If no such $D$ exists we set
$c_1(X,d)=\infty$). $(X,d)$ is said to be a metric space of negative
type, or a squared $L_2$ metric space, if the metric space
$\left(X,\sqrt{d}\right)$ admits an isometric embedding into Hilbert
space. A key example of a metric space of negative type is the
Banach space $L_1$. The purpose of this paper is to prove the
following result:

\begin{theorem}\label{thm:main}
For every $n\in \N$ there exists an $n$-point metric space $(X,d)$
of negative type such that
$$
c_1(X,d)\ge (\log n)^c,
$$
where $c>0$ is a universal constant which can be explicitly
estimated (see Section~\ref{sec:ex}).
\end{theorem}
The previous best known lower bound  in the setting of
Theorem~\ref{thm:main} is $c_1(X,d)=\Omega(\log \log n)$: this is
proved in~\cite{KR06} as an improved analysis of the spaces
constructed in the breakthrough result of~\cite{KV04}. The best
known upper bound~\cite{ALN08} for the $L_1$ distortion of finite
metric spaces of negative type is $c_1(X,d)=O\left(\left(\log
n\right)^{\frac12 +o(1)}\right)$, improving the previously known
bounds of $O\left((\log n)^{\frac34}\right)$ from~\cite{CGR08} and
the earlier bound of $O(\log n)$ from~\cite{Bou85} which holds for
arbitrary $n$-point metric spaces, i.e., without assuming negative
type.

Next we discuss the significance of Theorem~\ref{thm:main} in the
context of approximation algorithms.
The Sparsest Cut problem with general demands is a fundamental
combinatorial optimization problem which is defined as follows.
Given $n\in \N$ and two symmetric functions
$$C,D:\{1,\ldots,n\}\times\{1,\ldots,n\}\to [0,\infty)$$ (called
capacities and demands, respectively) and a subset $S \subseteq
\{1,\ldots,n\}$, write
$$
\Phi(S) \coloneqq \frac{\sum_{i=1}^n\sum_{j=1}^n C(i,j) \cdot |{\bf
1}_S(i) - {\bf 1}_S(j)|}
               {\sum_{i=1}^n\sum_{j=1}^n D(i,j) \cdot |{\bf 1}_S(i) - {\bf 1}_S(j)|},
$$
where ${\bf 1}_S$ is the characteristic function of $S$. The value
$$\Phi^*(C,D) \coloneqq \min_{S \subseteq \{1,\ldots,n\}} \Phi(S)$$ is
the minimum over all cuts (partitions) of $\{1,\ldots,n\}$ of the
ratio between the total capacity crossing the boundary of the cut
and the total demand crossing the boundary of the cut.

Finding in
polynomial time a cut for which $\Phi^*(C,D)$ is attained up to a
definite multiplicative constant is called the Sparsest Cut problem,
which is a basic step in approximation algorithms for several
NP-hard problems~\cite{LR99,AKRR90,Shmoys95,Chawla08}. Computing
$\Phi^*(C,D)$ exactly has been long-known to be NP-hard~\cite{SM90}.
More recently, it was shown in~\cite{CK07} that there exists
$\e_0>0$ such that it is NP-hard to approximate $\Phi^*(C,D)$ to
within a factor smaller than $1+\e_0$. In~\cite{KV04,CKKRS06} it was
shown that it is Unique Games hard to approximate $\Phi^*(C,D)$ to
within any constant factor (see~\cite{Khot02} for more information
on the Unique Games Conjecture).

The Sparsest Cut problem is the first algorithmic problem for which
bi-Lipschitz embeddings of metric spaces were successfully used to
design non-trivial polynomial time approximation
algorithms~\cite{LLR95,AR98}. While early results were based on a
remarkable approach using linear programming, an improved approach
based on semidefinite programming (SDP) was put forth by Goemans and
Linial in the late 1990s (see~\cite{Goe97,Lin02}). This approach
yields the best known approximation algorithm to the Sparsest Cut
problem~\cite{ALN08}, which has an approximation guarantee of
$O\left(\left(\log n\right)^{\frac12 +o(1)}\right)$. The SDP
approach of Goemans and Linial is based on computing the following
value:
\begin{multline}\label{eq:SDP}
M^*(C,D)\coloneqq \min\left\{\frac{\sum_{i=1}^n\sum_{j=1}^n
C(i,j)d(i,j)}{\sum_{i=1}^n\sum_{j=1}^n D(i,j)d(i,j)}:\right.\\
\left.\phantom{\frac{\sum_{i=1}^n\sum_{j=1}^n
C(i,j)d(i,j)}{\sum_{i=1}^n\sum_{j=1}^n D(i,j)d(i,j)}}\!\!\!\!\!\!\!\!\!\!\!\!\!\!\!\!\!\!\!\!\!\!\!\!\!\!\!\!\!\!\!\!\!\!\!\!\!\!\!\!\!\!\!\!\!\!\!\!\!\!(\{1,\ldots,n\},d)\ \mathrm{is\  a\  metric\  space\  of\  negative\
type}\right\}.
\end{multline}
The minimization problem in~\eqref{eq:SDP} can be cast as a
semidefinite program, and hence can be solved in polynomial time
with arbitrarily good precision (see the explanation
in~\cite{ALN08}). It is also trivial to check that $M^*(C,D)\le
\Phi^*(C,D)$, i.e., \eqref{eq:SDP} is a {\em relaxation} of the problem of
computing $\Phi^*(C,D)$. The {\em integrality gap} of this SDP is
the supremum of $\frac{\Phi^*(C,D)}{M^*(C,D)}$ over all symmetric
functions $C,D:\{1,\ldots,n\}\times\{1,\ldots,n\}\to [0,\infty)$.

The integrality gap of the Goemans-Linial SDP is well known to equal
the largest $L_1$ distortion of an $n$-point metric space of
negative type in $L_1$. We recall the argument. The cut cone
representation of $L_1$ metrics~\cite{dezalaur} states that a finite
metric space $(X,d)$ is isometric to a subset of $L_1$ if and only
if it is possible to associate to every subset $S\subseteq X$ a
non-negative number $\lambda_S\ge 0$ such that the distance between
any two points $x,y\in X$ can be computed via the formula:
$$d(x,y)=\sum_{S\subseteq X} \lambda_S|\1_S(x)-\1_S(y)|.$$ This fact
immediately implies that for all symmetric functions
$C,D:\{1,\ldots,n\}\times\{1,\ldots,n\}\to [0,\infty)$ we have:
\begin{equation}\label{eq:L_1 appears}
\Phi^*(C,D)=\min_{x_1,\ldots,x_n\in
L_1}\frac{\sum_{i=1}^n\sum_{j=1}^n C(i,j) \cdot \|x_i-x_j\|_1}
               {\sum_{i=1}^n\sum_{j=1}^n D(i,j) \cdot\|x_i-x_j\|_1}.
\end{equation}
Thus, for all $C,D$ and every metric $d$ on $\{1,\ldots,n\}$ we have:
\begin{equation}\label{eq:upper gap}
\frac{\sum_{i=1}^n\sum_{j=1}^n
C(i,j)d(i,j)}{\sum_{i=1}^n\sum_{j=1}^n D(i,j)d(i,j)}\ge
\frac{\Phi^*(C,D)}{c_1(\{1,\ldots,n\},d)}.
\end{equation}
Relation~\eqref{eq:L_1 appears} and the bound~\eqref{eq:upper gap}
explain how the result of~\cite{ALN08} quoted above yields an
algorithm for Sparsest Cut with approximation guarantee of
$O\left(\left(\log n\right)^{\frac12 +o(1)}\right)$.

In the reverse direction, given any metric $d$ on $\{1,\ldots,n\}$,
by a duality argument (see Proposition 15.5.2 and Exercise 4 in
chapter 15 of~\cite{Mat01}) there exist symmetric functions
$C_d,D_d:\{1,\ldots,n\}\times\{1,\ldots,n\}\to [0,\infty)$ (which
arise in~\cite{Mat01} from an appropriate separating hyperplane
between certain convex cones) satisfying for every $x_1,\ldots,x_n\in L_1$:
\begin{eqnarray}\label{eq:cheeger}\sum_{i=1}^n\sum_{j=1}^n C_d(i,j) \cdot \|x_i-x_j\|_1\ge
\sum_{i=1}^n\sum_{j=1}^n D_d(i,j) \cdot\|x_i-x_j\|_1,
\end{eqnarray}
and
\begin{eqnarray}\label{eq:no cheeger in d}\frac{\sum_{i=1}^n\sum_{j=1}^n C_d(i,j)d(i,j)}{\sum_{i=1}^n\sum_{j=1}^n D_d(i,j)d(i,j)}\le
\frac{1}{c_1(\{1,\ldots,n\},d)}.
\end{eqnarray}
A combination of~\eqref{eq:L_1 appears} and~\eqref{eq:cheeger} shows
that $\Phi^*(C_d,D_d)\ge 1$. Hence, choosing $C=C_d$ and $D=D_d$
in~\eqref{eq:upper gap}, together with~\eqref{eq:no cheeger in d},
implies that actually $\Phi^*(C_d,D_d)= 1$ and:
\begin{equation}\label{eq:M bound}
\frac{\sum_{i=1}^n\sum_{j=1}^n
C_d(i,j)d(i,j)}{\sum_{i=1}^n\sum_{j=1}^n D_d(i,j)d(i,j)}=\frac{\Phi^*(C_d,D_d)}{c_1(\{1,\ldots,n\},d)}.
\end{equation}
Substituting the metric $d$ from Theorem~\ref{thm:main}
into~\eqref{eq:M bound} yields the following theorem:
\begin{theorem}\label{thm:gap coro}
For every $n\in \N$ there exist symmetric functions
$C,D:\{1,\ldots,n\}\times\{1,\ldots,n\}\to [0,\infty)$ such that
$$
\frac{\Phi^*(C,D)}{M^*(C,D)}\ge (\log n)^c,
$$
where $c>0$ is the constant from Theorem~\ref{thm:main}. Thus, the integrality gap of the Goemans-Linial SDP for Sparsest
Cut is $(\log n)^{\Omega(1)}$.
\end{theorem}

\begin{remark} {\em The Sparsest Cut problem has an important special case called the Uniform Sparsest Cut problem (or also Sparsest Cut with uniform demands). This problem corresponds to the case where $C(i,j)\in \{0,1\}$ and $D(i,j)=1$ for all $i,j\in \{1,\ldots,n\}$. In this case $C$ induces a graph structure $G$ on $V=\{1,\ldots,n\}$, where two distinct $i,j\in V$ are joined by an edge if and only if $C(i,j)=1$. Thus for $S\subseteq V$ we have that $\Phi(S)$ is the number of edges joining $S$ and $V\setminus S$ divided by $|S|(n-|S|)$, and hence $n\Phi^*(S)$ is, up to a factor of $2$, the {\em edge expansion} of the  graph $G$.

The best known approximation algorithm for the Uniform Sparsest Cut problem~\cite{ARV04} achieves an approximation ratio of $O\left(\sqrt{\log n}\right)$, improving upon the previously best known bound~\cite{LR99} of $O(\log n)$.  The $O\left(\sqrt{\log n}\right)$ approximation algorithm of~\cite{ARV04} also uses the Goemans-Linial SDP relaxation described above. The best known lower bound~\cite{DKSV06} on the integrality gap of the Goemans-Linial SDP relaxation in the case of uniform demands  is $\Omega(\log \log n)$.

Our integrality gap example in Theorem~\ref{thm:main} works for the case of general demands, but {\em cannot} yield a lower bound tending to $\infty$ in the case of uniform demands, for the following reason. An inspection of the above argument shows that the integrality gap of the Goemans-Linial SDP in the case of uniform demands corresponds to the worst {\em average distortion}  of negative type metrics $d$ on $\{1,\dots, n\}$ into $L_1$, i.e., the infimum over $D>0$ such that for all negative type metrics $d$ on $\{1,\dots, n\}$ there exists a mapping $f:\{1,\ldots,n\}\to L_1$ for which
$$\|f(i)-f(j)\|_1\le Dd(x,y) \quad \forall i,j\in \{1,\ldots,n\},$$ and $$\sum_{i,j=1}^n \|f(i)-f(j)\|_1\ge \sum_{i,j=1}^n d(i,j).$$ This connection between the Uniform Sparsest Cut problem and average distortion embeddings is explained in detail in~\cite{Rab08}. The metric spaces in Theorem~\ref{thm:main} have doubling constant $O(1)$, 
and therefore by the proof in~\cite{Rab08} they admit an embedding into the real line (and hence also into $L_1$) with average distortion $O(1)$\footnote{In~\cite{Rab08} this fact is not explicitly stated for doubling metrics, but the proof only uses the so called ``padded decomposability" of the metric $d$ (see~\cite{KLMN05} for a discussion of this notion), and it is a classical fact (which is implicit in~\cite{Ass83}) that doubling metric spaces satisfy this property.}. Thus our work does not provide progress on the problem of estimating the asymptotic behavior of the integrality gap of the SDP for Uniform Sparsest Cut, and it remains an interesting open problem to determine whether the currently best known lower bound, which is $\Omega(\log \log n)$, can be improved to $(\log n)^{\Omega(1)}$.  }
\end{remark}



\section{The example}\label{sec:ex}
Define $\rho:\R^3\times \R^3\to [0,\infty)$ by
\begin{eqnarray}\label{eq:def rho}
&&\!\!\!\!\!\!\!\!\!\!\!\!\!\!\!\!\!\!\!\nonumber\rho\Big( (x,y,z), (t,u,v) \Big)\\
 &\coloneqq& \left(\left[  \left((t-x)^2 +
 (u-y)^2\right)^2 + (v-z+2xu - 2yt)^2\right]^{\frac{1}{2}}\right. \nonumber \\&\phantom{\le}&+ (t-x)^2 +
(u-y)^2\Bigg)^{\frac{1}{2}}.
\end{eqnarray}

It was shown in~\cite{LN06} that $(\R^3,\rho)$ is a metric space of
negative type. The result of~\cite{CK06} gives
$c_1(\R^3,\rho)=\infty$, which implies that
$c_1\left(\{0,1,\ldots,k\}^3,\rho\right)$ tends to $\infty$ with $k$
(The proof of this implication is via a compactness argument which
would fail if $c_1(\R^3,\rho)$ were defined using the sequence space
$\ell_1$ rather than the function space $L_1$).
Theorem~\ref{thm:main} follows from a quantitative refinement of the
statement $c_1(\R^3,\rho)=\infty$:
\begin{theorem}\label{thm:grid}
There exist universal constants $\psi,\delta>0$ such that for all
$k\in \N$ we have:
$$
c_1\left(\{0,1,\ldots,k\}^3,\rho\right)\ge \psi(\log k)^\delta.
$$
\end{theorem}

The proof of Theorem~\ref{thm:grid} is quite lengthy and involved.
Complete details are given in the forthcoming full version of this paper~\cite{ckn}.
Here
we will give the key concepts and steps in the proof. First we wish
to highlight a natural concrete open question that arises from
Theorem~\ref{thm:grid}. Denote:
$$
\delta^*\coloneqq \limsup_{k\to\infty}
\frac{\log\left(c_1\left(\{0,1,\ldots,k\}^3,\rho\right)\right)}{\log\log
k}.
$$
Combining the result of~\cite{ALN08} and Theorem~\ref{thm:grid}
shows that $\delta^*\in[\delta,1/2]$ for some universal constant
$\delta>0$. In~\cite{ckn} we will give an explicit (though non-sharp) lower estimate on $\delta$ (just for the sake of stating a concrete bound in this paper, we can safely assert at this juncture that, say,
 $\delta \ge2^{-1000}$).
  Proposition 7.10 in~\cite{ckn} (which we need to iterate 6 times)
 is the most involved step and essentially the only place
in which sharpness has been sacrificed to simplify the exposition.
We do not know how close an optimal version of our argument would
come to yielding the constant
 $\delta^*$ . Conceivably $\delta^*= \frac12$. If so, the metric
 spaces from Theorem~\ref{thm:grid} will already show that
the integrality gap of the Sparsest Cut SDP is $\Theta\left((\log
n)^{\frac12 +o(1)}\right)$.

\section{Quantitative central collapse}
The main result of~\cite{CK06} states that if $U\subseteq \R^3$ is
an open subset and if $f:U\to L_1$ is a Lipschitz function in the
metric $\rho$ defined in~\eqref{eq:def rho} then for almost every
(with respect to Lebesgue measure) $(x,y,z)\in U$ we have
\begin{equation}\label{eq:collapse}
\lim_{\e\to 0^+}
\frac{\|f(x,y,z+\e)-f(x,y,z)\|_1}{\rho\big((x,y,z+\e),(x,y,z)\big)}=0.
\end{equation}
Our main result is the following quantitative version of this
statement:
\begin{theorem}\label{thm:rate}
There exists a universal constant $\delta\in (0,1)$ with the
following property. Let $B\subseteq \R^3$ be a unit ball in the
metric $\rho$ and let $f:B\to L_1$ be a function which is
$1$-Lipschitz with respect to $\rho$. Then for every $\e\in (0,1/4)$
there exists $r\ge \e$ and $(x,y,z)\in B$ such that $(x,y,z+r)\in B$ and:
\begin{equation*}
\frac{\|f(x,y,z+r)-f(x,y,z)\|_1}{\rho\big((x,y,z+r),(x,y,z)\big)}\le
\frac{1}{\left(\log(1/\e)\right)^\delta}.
\end{equation*}
\end{theorem}
 It was shown in Remark 1.6 of~\cite{LN06}
that Theorem~\ref{thm:rate} (which was not known at the time)
implies that if $X\subseteq \R^3$ is an $\eta$-net in the unit ball
with respect to $\rho$ centered at $(0,0,0)$ for some $\eta\in
(0,1/16)$ then $c_1(X,\rho)=\Omega(1)(\log(1/\eta))^\delta$. The key
point of~\cite{LN06} is that one can use a Lipschitz extension
theorem for doubling metric spaces~\cite{LN05} to extend an
embedding of $X$ into $L_1$ to a Lipschitz (but not
bi-Lipschitz) function defined on all of $\R^3$ while incurring a
universal multiplicative loss in the Lipschitz constant (in fact, since we are extending from a net,
the existence of the required Lipschitz extension also follows from a simple partition of
unity argument and there is no need to use the general result of~\cite{LN05}). Since the
collapse result in Theorem~\ref{thm:rate} for this extended function
occurs at a definite scale, one can use the fact that the function
is bi-Lipschitz on the net $X$ to obtain the required lower bound on
the distortion. The metric space $
\left(\frac{\left\{0,\ldots,k\right\}}{k}\times
\frac{\left\{0,\ldots,k\right\}}{k}\times\frac{\left\{0,\ldots,k\right\}}{k^2},\rho\right)
$ is isometric to the metric space
$\left(\{0,1,\ldots,k\}^3,\frac{\rho}{k}\right)$, and it contains
such an $\eta$ net $X$ with $\eta\approx \frac{1}{k}$. Hence
Theorem~\ref{thm:rate} in conjunction with the above discussion
implies Theorem~\ref{thm:grid}.

In the remainder of this extended abstract we will explain the
ingredients that go into the proof of Theorem~\ref{thm:rate}.

\section{The Heisenberg group}

Equip $\R^3$ with the following group structure: $$ (a,b,c)\cdot
(\alpha,\beta,\gamma)\coloneqq
(a+\alpha,b+\beta,c+\gamma+a\beta-b\alpha).$$ The resulting
non-commutative group is called the Heisenberg group, and is denoted
$\H$. Note that the identity element of $\H$ is $e=(0,0,0)$ and the
inverse of $(a,b,c)\in \H$ is $(-a,-b,-c)$. The center of $\H$ is
$\{0\}\times\{0\}\times \R$. This explains why we call results such
as~\eqref{eq:collapse} ``central collapse".

For every $g=(a,b,c)\in \H$ we associate a special affine $2$-plane,
called the horizontal $2$-plane at $g$, which is defined as
$H_g=g\cdot\left(\R^2\times \{0\}\right)$. Thus $H_e$ is simply the
$x,y$ plane. The Carnot-Caratheodory metric  on $\H$, denoted
$d^\H$, is defined as follows: for $g,h\in \H$, $d^\H(g,h)$ is the
infimum of lengths of smooth curves $\gamma:[0,1]\to \H$ such that
$\gamma(0)=g$, $\gamma(1)=h$ and for all $t\in [0,1]$ we have
$\gamma'(t)\in H_{\gamma(t)}$ (i.e., the tangent vector at time $t$
is restricted to be in the corresponding horizontal $2$-plane. The standard Euclidean norm on $\R^2$ induces a natural Euclidean norm on $H_g$ for all $g\in \H$, and hence the norm of $\gamma'(t)$ is well defined for all $t\in [0,1]$. This is how the length of $\gamma$ is computed). For
concreteness we mention that the metric $d^\H$ restricted to the
integer grid $\mathbb Z^3$ is bi-Lipschitz equivalent to the word
metric on $\H$ induced by the following (and hence any finite)
canonical set of generators: $\{(\pm1,0,0),(0,\pm1,0),(0,0,\pm1)\}$ (in other words, this is simply the shortest path metric on the Cayley graph given by these
  generators).
The metric space $(\H,d^\H)$ is bi-Lipschitz equivalent to
$(\R^3,\rho)$ via the mapping $(x,y,z)\mapsto
\left(\frac{x}{\sqrt{2}},\frac{y}{\sqrt{2}},z\right)$ (this follows
from the ``ball-box theorem"---see for
example~\cite{Mont02})
. Hence in what follows it
will suffice to prove Theorem~\ref{thm:rate} with the metric $\rho$
replaced by the metric $d^\H$.

Below, for $r>0$ and $x\in \H$, we denote by $B_r(x)$ the open ball in the
metric $d^\H$ of radius $r$ centered at $x$. The
following terminology will be used throughout this paper. A half
space in $\H$ is the set of points lying on one side of some affine
$2$-plane in $\R^3$, including the points of the plane itself. A half space is
called horizontal if its associated $2$-plane is of the form $H_g$
for some $g\in \H$. Otherwise the half space is called vertical. An
 affine line in $\R^3$ which passes through some point $g\in
\H$ and lies in the plane $H_g$ is called a horizontal line. The set
of all horizontal lines in $\H$ is denoted $\lines(\H)$.

\section{Cut measures and sets of finite perimeter}\label{sec:cut}

In what follows we set $B=B_1(e)=B_1((0,0,0))$ and fix a $1$-Lipschitz
function $f:B\to L_1$ (in the metric $d^\H$). The cut (semi)-metric
associated to a subset $E\subseteq B$ is defined as
$d_E(x,y)\coloneqq |\1_E(x)-\1_E(y)|$. Let $\cut(B)$ denote the
space of all measurable cuts (subsets) of $B$ equipped with the
semi-metric given by the Lebesgue measure of the symmetric
difference. In~\cite{CK06} a measure theoretic version of the cut-cone
representation was studied. It states that there is a canonical
Borel measure $\Sigma_f$ on $\cut(B)$ such that for almost all
$x,y\in B$ we have:
\begin{equation}\label{eq:representation}
d_f(x,y)\coloneqq
\|f(x)-f(y)\|_1=\int_{\cut(B)}d_E(x,y)d\Sigma_f(E).
\end{equation}
A key new ingredient of the result of~\cite{CK06} is that the
Lipschitz condition on $f$ forces the measure $\Sigma_f$ to be
supported on cuts with additional structure, namely cuts with finite
perimeter. For sets with smooth boundary the perimeter is a certain
explicit integral with respect to the surface area measure on the
boundary (and, in the case of $\R^3$ equipped with the Euclidean
metric, it simply coincides with the surface area for smooth sets).
However, since the sets appearing in the
representation~\eqref{eq:representation} cannot be a priori enforced
to have any smoothness properties we need to work with a measure
theoretical extension of the notion of surface area. Namely, define
for every $E\in \cut (B)$, and an open set $U\subseteq \H$
\begin{multline}\label{eq:def per}
\per(E)(U)\coloneqq \inf\left\{\liminf_{i\to \infty}
\int_U\Lip_x(h_i)d\mu(x): \{h_i\}_{i=1}^\infty\ \right.\\ \mathrm{Lipschitz\
functions\  tending\  to\ } \1_E\  \mathrm{in}\ L_1^{{\rm loc}}(B)
\Bigg\}.
\end{multline}
Here, and in what follows, $\mu$ denotes the Lebesgue measure on
$\H=\R^3$ and for $h:\H\to \R$ the quantity
$$\Lip_x(h)\coloneqq\limsup_{y\to x}\frac{|h(y)-h(x)|}{d^\H(x,y)}$$
denotes the local Lipschitz constant of $h$ at $x$. Convergence in $L_1^{{\rm loc}}(B)$ means, as usual, convergence in $L_1(K,\mu)$ for all compact subsets $K\subseteq B$. (To get some intuition for this notion, consider the analogous definition in the Euclidean space $\R^3$, i.e., when the functions $\{h_i\}_{i=1}^\infty$ are assumed to be Lipschitz with respect to the Euclidean metric rather than the metric $d^\H$. In this case, for sets $E$ with smooth boundary, the quantity $\per(E)(U)$ is the surface area of the part of the boundary of $E$ which is contained in $U$).
$\per(E)(\cdot)$ can be extended to be a Radon measure on $\H$ (see
for example~\cite{luigetal}).
A key insight of~\cite{CK06} is that the fact that $f$ is
$1$-Lipschitz 
implies that for every open subset $U\subseteq B$ we have:
\begin{equation}\label{eq:total perimeter}
\int_{\cut(B)}\per(E)(U)d\Sigma_f(E)\le C\cdot \mu(U),
\end{equation}
where $C$ is a universal constant (independent of $f$). Also there
is an induced total perimeter measure $\lambda_f$ defined by:
\begin{equation}\label{eq:def lambda}
\lambda_f(\cdot)\coloneqq \int_{\cut(B)} \per(E)(\cdot)d\Sigma_f(E).
\end{equation}

In~\cite{CK06} the inequality~\eqref{eq:total perimeter} was used to
show that $\H$ does not admit a bi-Lipschitz embedding into $L_1$ by
exploiting the  infinitesimal regularity of sets of finite
perimeter. Specifically, let $E\subseteq \H$ be a set with finite
perimeter. Then, as proved in~\cite{italians1,italians2}, with
respect to the measure $\per(E)$, for almost every $p\in E$,
asymptotically under blow up the measure of the symmetric difference
of $E$ and some unique {\em vertical} half space goes to $0$.
Intuitively, this means that (in a measure theoretic sense) almost
every point $p\in
\partial E$ has a tangent $2$-plane which is vertical. Observe that a cut
semi-metric associated to a vertical half-space, when restricted
to a coset of the center of $\H$, is identically $0$. This fact
together with~\eqref{eq:representation} suggests that under blow-up,
at almost all points,  $f$ becomes degenerate in the direction of
cosets of the center, and therefore $\H$ does not admit a
bi-Lipschitz embedding into $L_1$. This is the heuristic argument
behind the main result of~\cite{CK06}. What is actually required is
a version of the results of~\cite{italians1,italians2} for measured
families of finite perimeter cuts corresponding to the
representation~\eqref{eq:representation}.

The verticality, which played a key role above, is an initially
surprising feature of the Heisenberg geometry, which in actuality,
can easily be made intuitively plausible. We will not do so here
since below we do not use it. What we do use is a quantitative
version of a cruder statement, which in effect ignores the issues of
verticality and uniqueness of generalized tangent planes. This
suffices for our purposes. Our approach incorporates ideas from a
second and simpler proof of the (non-quantitative) bi-Lipschitz
non-embeddability of $\H$ into $L_1$, which was obtained
in~\cite{ckmetmon}.
The second proof, which did not require the
results of~\cite{italians1,italians2}, is based on the notion of
{\em monotone sets} which we now describe.

\remove{ In order to use the above approach to prove a quantitative
result such as Theorem~\ref{thm:rate} one needs to obtain
quantitative versions of the blow-up results
of~\cite{italians1,italians2} (and the corresponding earlier results
in~\cite{luigi1,luigi2} on which they depend). While it might be
possible to accomplish this with enough effort, we do not know how
to do so or whether the resulting estimates will suffice for our
purposes. We therefore take a different route based on an
alternative simpler proof of the bi-Lipschitz non-embeddability of
$\H$ into $L_1$ which was obtained in~\cite{ckmetmon} (which in the
end leads to a quantitative version of a result which is weaker than
the result of~\cite{italians1,italians2} but is still sufficient for
our purposes). This approach is based on the notion of monotone sets
which we now describe.}

\section{Monotone sets}\label{sec:monotone}

Fix an open set $U\subseteq \H$. Let $\lines(U)$ denote the space of
unparametrized oriented horizontal lines whose intersection with $U$
is nonempty. Let $\mathcal N_U$ denote the unique left invariant
measure on $\lines(\H)$ normalized so that $\mathcal
N_U(\lines(U))=1$. A subset $E\subseteq U$ is {\em monotone with
respect to $U$} if for $\mathcal N_U$-almost every line $L$, both $E\cap L$ and $(U\setminus E)\cap L$ are essentially connected, in the sense that there exist connected subsets $F_L=F_L(E),F_L'=F_L'(E)\subseteq L$ (i.e., each of $F_L,F_L'$ is either
empty, equals $L$, or is an interval, or a ray in $L$) such that the
symmetric differences $(E\cap L)\triangle F_L$ and $((U\setminus
E)\cap L)\triangle F_L'$ have $1$-dimensional Hausdorff measure $0$.

When $U=\H$, a non-trivial classification theorem was proved
in~\cite{ckmetmon}, stating that if $E$ is monotone with respect to
$\H$ then either $E$ or $\H\setminus E$ has measure zero, or there
exists a half space $\mathcal P$ such that $\mu(E\triangle \mathcal
P)=0$. Note for the sake of comparison with the Euclidean case that
if we drop the requirement that the lines are horizontal in the
definition of monotone sets then monotonicity would essentially mean
that (up to sets of measure $0$) both $E$ and the complement of $E$
are convex sets, and hence $E$ is a half space up to a set of
measure $0$. The non-trivial point in the classification result
of~\cite{ckmetmon} is that we are allowed to work only with a
codimension $1$ subset of all affine lines in $\R^3$, namely the
horizontal lines.

Using the above classification result for monotone sets,
in~\cite{ckmetmon} the non-embedding result for $\H$ in $L_1$ is
proved by using once more a blow-up argument (or metric differentiation) to reduce the
non-embedding theorem to the special case in which the cut measure
$\Sigma_f$ is supported on sets which are monotone with respect to
$\H$. Thus, the cut measure is actually supported on half spaces. It
follows (after the fact) that the connectedness condition in the
definition of monotone sets holds for every line $L$, not just for
horizontal lines. This implies that for {\em every  affine} line
$L$, if $x_1,x_2,x_3\in L$ and $x_2$ lies between $x_1$ and $x_3$
then
\begin{equation}\label{additivity}
\|f(x_1)-f(x_3)\|_1=\|f(x_1)-f(x_2)\|_1+\|f(x_2)-f(x_3)\|_1.
\end{equation}
 But if $L$ is vertical then $d^\H|_L$ is bi-Lipschitz to
the {\em square root} of the difference of the $z$-coordinates, and
it is trivial to verify that this metric on $L$ is not bi-Lipschitz
equivalent to a metric on $L$ satisfying~\eqref{additivity}.


In proving Theorem~\ref{thm:rate}, the most difficult part by far is
a stability theorem stating in quantitative form that individual
cuts which are ``approximately monotone" are close to half spaces;
see Theorem~\ref{thm:stability}. Here, it is important to have the
right notion of ``approximately monotone". We also show that on a
controlled scale, modulo a controlled error, we can at most
locations  reduce to the case when the cut measure is supported on
cuts which are approximately close to being monotone so that
Theorem~\ref{thm:stability} can be applied, and such that in
addition there is a bound on the total cut measure. For this, the
bound~\eqref{eq:total perimeter} is crucially used to estimate the
scale at which the ``total non-monotonicity" is appropriately small.
At such a good scale and location, it now follows that up to a small
controlled error~\eqref{additivity} holds. In the next section we
introduce the notion of $\delta$-monotone sets and state the
stability theorem which ensures that $\delta$-monotone sets are
close to half spaces on a ball of controlled size.

\remove{ Our proof of Theorem~\ref{thm:rate} is based on a
quantitative version of the above sketch. The heart of the matter is
to show that on a controlled scale, cuts which are ``approximately
monotone" are close to half spaces---the key issue here is an
appropriate definition of ``approximately monotone", and an argument
showing that modulo a controlled error we can reduce to the case
when the cut measure is supported on cuts which are appropriately
close to being monotone (it is in the latter statement that the
bound~\eqref{eq:total perimeter} is crucially used to estimate the
scale at which the total non-monotonicity is so small so that the
suggested scheme applies). In the next section we introduce the
notion of $\delta$-monotone sets and state our stability result
which ensures that $\delta$-monotone sets are close to half spaces
on a ball of controlled size.}

\section{Stability of monotone sets}

Denote $\mathcal N=\mathcal N_{B}$, i.e., $\mathcal N$ is the left
invariant measure on $\lines (\H)$ normalized so that the measure of
the horizontal lines that intersect $B$ is $1$. For a horizontal
line $L\in \lines(\H)$ let $\mathcal H_L^1$ denote the
$1$-dimensional Hausdorff measure on $L$ with respect to the metric
induced from $d^\H$.

Fix a ball $B_r(x)\subseteq B$. For every measurable $E\subseteq \H$
and $L\in \lines(B_r(x))$ we define the non-convexity of $(E,L)$ on
$B_r(x)$ by:
\begin{multline}\label{eq:NC}
{\rm NC}_{B_r(x)}(E,L)\coloneqq \inf\left\{\int_{L\cap
B_r(x)}\left|\1_I-\1_{E\cap L\cap B_r(x)}\right|d\mathcal H_L^1:\ \right.\\
I\subseteq L\cap B_r(x)\ \mathrm{subinterval} \Bigg\}.
\end{multline}
The
non-monotonicity of $(E,L)$ on $B_r(x)$ is defined as:
$$
{\rm NM}_{B_r(x)}(E,L)\coloneqq {\rm NC}_{B_r(x)}(E,L)+{\rm NC}_{B_r(x)}(\H\setminus
E,L).
$$
The non-monotonicity of $E$ on $B_r(x)$ is defined as:
\begin{eqnarray*}
&&\!\!\!\!\!\!\!\!\!\!\!\!\!\!\!{\rm NM}_{B_r(x)}(E)\coloneqq \frac{1}{r^4}\int_{\lines(B_r(x))}{\rm
NM}_{B_r(x)}(E,L)d\mathcal{N}(L)\\ &=&
\frac{1}{\mathcal{N}(\lines(B_r(x)))}\int_{\lines(B_r(x))}\frac{{\rm
NM}_{B_r(x)}(E,L)}{r}d\mathcal{N}(L).
\end{eqnarray*}
Note that by design  ${\rm NM}_{B_r(x)}(E)$ is a scale invariant
quantity. A measurable set $E$ is said to be $\delta$-monotone on
$B_r(x)$ if ${\rm NM}_{B_r(x)}(E)<\delta$. Our stability result for
monotone sets is the following theorem:
\begin{theorem}\label{thm:stability}
There exists a universal constant $a>0$ such that if a measurable
set $E\subseteq B_r(x)$ is $\e^a$-monotone on $B_r(x)$ then there
exists a half-space $\mathcal P$ such that
$$
\frac{\mu\left((E\cap B_{\e r}(x))\triangle \mathcal
P\right)}{\mu(B_{\e r}(x))}<\e^{1/3}.
$$
\end{theorem}

The proof of Theorem~\ref{thm:stability} constitutes the bulk of the
full version of this paper~\cite{ckn}. Formally, it follows the steps of the
argument of~\cite{ckmetmon} in the case of sets which are precisely
monotone. However, substantial additions are required arising from
the need to work with certain appropriate quantitatively defined
notions of ``fuzzy"  measure theoretical boundaries of sets, and by
the need to make a certain existence statement of~\cite{ckmetmon}
quantitative. 

\section{Splitting the cut measure}\label{sec:split}

Theorem~\ref{thm:stability} will allow us to control individual
integrands in the cut representation~\eqref{eq:representation}
(assuming that we can find a scale at which the total
non-monotonicity is small enough---this is discussed in
Section~\ref{sec:choose monotone} below). But, such point-wise
estimates do not suffice since we do not have any a priori control
on the total mass of the cut measure $\Sigma_f$. To overcome this
problem we split the measure $\Sigma_f$ into two parts in such a way
that one part has controlled total mass, while the other part
contributes a negligible amount to the metric $d_f$.

Fix a ball $B_r(p)\subseteq B$.
In what follows we will use the notation $\lesssim,\gtrsim$ to
denote the corresponding inequalities up to universal factors. We
shall also use the fact that $\mu(B_s(z))=s^4\mu(B_1(z)$ for all $s>0$
and $z\in \H$.

For $\theta>0$ define $D_\theta\subseteq \cut(B)$ by
$$D_\theta\coloneqq \{E\in \cut(B):\
\per(E)(B_r(p))>\theta\mu(B_r(p))\}.$$ Markov's inequality combined
with~\eqref{eq:total perimeter} implies that
$\Sigma_f(D_\theta)\lesssim \frac{1}{\theta}$. Define a semi-metric
$d_\theta$ on $\H$ by $$d_\theta(x,y)\coloneqq
\int_{D_\theta}d_E(x,y)d\Sigma_f(E).$$ We claim that even though we
do not have a bound on $\Sigma_f(\cut(B))$ we can still control the
distance between $d_f$ and $d_\theta$ in $L_1(B_r(p)\times B_r(p))$.
This can be deduced from  the  isoperimetric inequality on $\H$
(see~\cite{CDF94}) which implies that for every $E\in \cut(B)$ we
have
\begin{equation*}
\frac{\mu(B_r(p)\cap E)}{\mu(B_r(p))}\cdot\frac{\mu(B_r(p)\setminus
E)}{\mu(B_r(p))}\lesssim
\left(\frac{r}{\mu(B_r(p))}\per(E)(B_r(p))\right)^{4/3},
\end{equation*}
or
\begin{equation}\label{eq:sobolev}
\mu(B_r(p)\cap E)\mu(B_r(p)\setminus E)\lesssim
r^4\left(\per(E)(B_r(p))\right)^{4/3}. \end{equation} The argument
is as follows:
 for each non-negative integer $n$
define \begin{multline*} A_n\coloneqq \Bigg\{E\in \cut(B):\\
\left.\frac{\theta\mu(B_r(p))}{2^{n+1}}<
\per(E)(B_r(p))\le\frac{\theta\mu(B_r(p))}{2^n}\right\}.\end{multline*} Then
$$\cut(B)\setminus D_\theta=\left(\bigcup_{n=0}^\infty
A_n\right)\bigcup A_\infty,$$ where $$A_\infty\coloneqq \{E\in
\cut(B):\ \per(E)(B_r(p))=0\}.$$ Markov's inequality combined
with~\eqref{eq:total perimeter} implies that $\Sigma_f(A_n)\lesssim
\frac{2^n}{\theta}$ for all $n$, while~\eqref{eq:sobolev} implies
that for each $E\in A_n$ we have $$\mu(B_r(p)\cap
E)\mu(B_r(p)\setminus E)\lesssim r^{28/3}
\left(\frac{\theta}{2^n}\right)^{4/3}$$ and for $E\in A_\infty$ we
have $\mu(B_r(p)\cap E)\mu(B_r(p)\setminus E)=0$.  We therefore
obtain the estimate:
\begin{eqnarray}\label{eq:before sobolev}
&&\!\!\!\!\!\!\!\!\!\!\!\nonumber\|d_f-d_\theta\|_{L_1(B_r(p)\times B_r(p))}\nonumber=\int_{B_r(p)\times
B_r(p)}\\&\phantom{\le}&\left(\int_{\cut(B_r(p))\setminus
D_\theta}|\1_E(x)-\1_E(y)|d\Sigma_f(E)\right) d\mu(x)d\mu(y)\nonumber\\ \nonumber
&=&\sum_{n=0}^\infty\int_{A_n}2\mu(B_r(p)\cap E)\mu(B_r(p)\setminus
E)d\Sigma_f(E)\\&\lesssim&\sum_{n=1}^\infty
\frac{2^n}{\theta}r^{28/3}\left(\frac{\theta}{2^n}\right)^{4/3}\lesssim
r^{28/3}\theta^{1/3}.
\end{eqnarray}

\section{Controlling the scale at which the total
non-monotonicity is small}\label{sec:choose monotone}

We shall require a formula, known as a kinematic formula, which
expresses the perimeter of a set $E\subseteq \H$ as an integral over
the space of lines $L$ of the perimeter of the $1$-dimensional sets
$E\cap L$. This formula (proved in Proposition 3.13
of~\cite{montefalcone}) asserts that there exists a constant
$\gamma=\gamma(\H)$ such that for every open subset $U\subseteq \H$
and a measurable subset $E\subseteq \H$ with $\per(E)(U)<\infty$ the
function $L\mapsto \per(E\cap L)(U\cap L)$ from $\lines(U)$ to
$[0,\infty)$ is in $L_1(\lines(U),\mathcal N)$ and satisfies the
identity:
\begin{equation}\label{eq:kinematic'}
\per(E)(U)=\gamma\int_{\lines(U)}\per(E\cap L)(U\cap L)d\mathcal
N(L).
\end{equation}
Here we used the notion of one dimensional perimeter, which is
defined analogously to~\eqref{eq:def per}. For one dimensional sets
with finite perimeter the notion of perimeter has a  simple
characterization (see Proposition 3.52 in~\cite{luigetal}). Whenever
$\per(E\cap L)(U\cap L)<\infty$ there exists a unique collection of
finitely many disjoint intervals $$\mathcal
I(E,L,U)=\{I_1(E,L,U),\ldots,I_n(E,L,U)\}$$ which are relatively
closed in $L\cap U$ and such that the symmetric difference of $E\cap
L$ and $\bigcup_{j=1}^nI_j(E,L,U)$ has measure $0$. The perimeter
measure $\per(E\cap L)$ is the sum of delta functions concentrated
at the end points of these intervals and hence $\per(E\cap L)(U\cap
L)$ is the number of these end points.

Fix  $\delta\in (0,1/2)$. For every non-negative integer $j$ let
$C_j(E,L)$ denote the collection of intervals in $\mathcal I(E,L,B)$
whose length is in $\left(2\delta^{j+1},2\delta^j\right]$. Let
$\E_j(E,L)$ denote the collection of all end points of intervals in
$C_j(E,L)$. For a measurable $A\subseteq B$ write:
\begin{multline}\label{eq:def wj(E)}
w_j(E)(A)\coloneqq\\ \gamma\int_{\lines(B)}\frac{|A\cap
\E_j(E,L)|+|A\cap \E_j(\H\setminus E,L)|}{2}d\mathcal N(L),
\end{multline}
where $\gamma$ is as in~\eqref{eq:kinematic'}. We also set:
\begin{equation}\label{eq:def wj}
w_j(A)=\int_{\cut(B)}w_j(E)(A)d\Sigma_f(E).
\end{equation}
The kinematic formula~\eqref{eq:kinematic'} implies that
\begin{equation}\label{eq:decompose w_j}
\lambda_f=\sum_{j=0}^\infty w_j.
\end{equation}
It follows from~\eqref{eq:decompose w_j} that \begin{equation}\label{eq:bound sum}\sum_{j=0}^\infty
w_j\left(B\right)=\lambda_f\left(B\right)\lesssim 1.\end{equation} Thus there
exists $j\le \delta^{-1}$ for which $w_j\left(B\right)\lesssim
\delta$. We shall fix this integer $j$ from now on. The ball $B$
contains $\gtrsim \delta^{-4j}$ disjoint balls of radius $\delta^j$.
Thus there exists $y\in B$ such that $B_{\delta^j}(y)\subseteq B$
and $w_j\left(B_{\delta^j}(y)\right)\lesssim \delta^{4j+1}$. We
shall fix this point $y\in B$ from now on.

Fix $E\subseteq \H$ with finite perimeter. For $\mathcal N$-almost
every $L\in \lines\left(B_{\delta^j(y)}\right)$ the set $\mathcal
I\left(E,L,B_{\delta^j}(y)\right)$ consists of finitely many
intervals $I_1,\ldots,I_n$. Note that each of the intervals
$I_1,\ldots,I_{n}$ (including both endpoints) is contained in the
closure of $B_{\delta^j}(y)$, and hence its length is at most
$2\delta^j$. It follows that each of these intervals lies in
$C_k(E,L)$ for some $k\ge j$. By the definition~\eqref{eq:NC}  we
have:
$$
{\rm NC}_{B_{\delta^j}(y)}(E,L)\le \sum_{s=1}^n\length(I_s)\lesssim
\sum_{k\ge j} \delta^k |B_{\delta^j}(y)\cap\E_k(E,L)|.
$$
Arguing similarly for $\H\setminus E$ yields:
\begin{multline}\label{eq:before integrate L}
{\rm NM}_{B_{\delta^j}(y)}(E,L)\lesssim \\\sum_{k\ge j} \delta^k
\left(|B_{\delta^j}(y)\cap\E_k(E,L)|+|B_{\delta^j}(y)\cap\E_k(\H\setminus
E,L)|\right).
\end{multline}
Averaging~\eqref{eq:before integrate L} over $L\in
\lines(B_{\delta^j}(y))$ gives a bound on the total
non-monotonicity:
\begin{eqnarray}\label{eq:before integrate sigma}
&&\!\!\!\!\!\!\!\!\!\!\!\!\!\nonumber{\rm NM}_{B_{\delta^j}(y)}(E)\lesssim \delta^{-4j}\sum_{k\ge
j}\delta^k\int_{\lines\left(B_{\delta^j}(y)\right)}\\\nonumber&\phantom{\le}&\left(|B_{\delta^j}(y)\cap\E_k(E,L)|+
|B_{\delta^j}(y)\cap\E_k(\H\setminus
E,L)|\right)d\mathcal N(L)\\
&\stackrel{\eqref{eq:def wj(E)}}{\lesssim}& \delta^{-4j}\sum_{k\ge
j}\delta^k w_k(E)\left(B_{\delta^j}(y)\right).
\end{eqnarray}
Integrating~\eqref{eq:before integrate sigma} with respect to $E\in
\cut(B)$ and using~\eqref{eq:def wj} yields the bound:
\begin{eqnarray}\label{eq:instead of lemma}
&&\!\!\!\!\!\!\!\!\!\!\!\!\!\!\!\nonumber\int_{\cut(B)}{\rm NM}_{B_{\delta^j}(y)}(E)d\Sigma_f(E)\lesssim
\delta^{-4j}\sum_{k\ge j} \delta^k
w_k\left(B_{\delta^j}(y)\right)\\\nonumber&\stackrel{\eqref{eq:decompose
w_j}}{\le}&\delta^{-3j}w_j\left(B_{\delta^j}(y)\right)+\delta^{-4j}
\cdot\delta^{j+1}\lambda_f\left(B_{\delta^j}(y)\right)\\\nonumber
&\stackrel{\eqref{eq:total perimeter}}{\lesssim}&
\delta^{-3j}w_j\left(B_{\delta^j}(y)\right)+\delta^{-3j+1}\mu\left(B_{\delta^j}(y)\right)\\&\lesssim&
\delta^{-3j}w_j\left(B_{\delta^j}(y)\right)+\delta^{j+1}\lesssim
\delta^{j+1},
\end{eqnarray}
where in the last inequality above we used our choice of $y$ and $j$
which ensures that $w_j\left(B_{\delta^j}(y)\right)\lesssim
\delta^{4j+1}$.


\renewcommand{\P}{\mathcal{P}}
\section{Cut metrics close to ones supported on almost half
spaces}\label{sec:half}

Let $\Sigma_\P$ be a measure on $\cut(B)$ which is supported on half
spaces. Assume that $$\|d_\P-d_f\|_{L_1(B\times B)}\le \e.$$ Our goal
is to use this assumption to deduce that $d_f$ must collapse some
pair of points lying on the same coset of the center whose distance
is controlled from below by an appropriate power of $\e$. Namely, we
will show that there exist $x,y\in B$ lying on the same coset of the center such that $$d^\H(x,y)\gtrsim
\e^{1/90}\quad\mathrm{yet}\quad d_f(x,y)\lesssim \e^{1/18}d^\H(x,y).$$ This step is a
quantitative (integral) version of the argument that was sketched in
Section~\ref{sec:monotone}, which relies on the fact that $d_\P$ is
additive along every affine line.

 Define $u=(0,0,h)$ where $h>0$ is a small enough
universal constant such that $\frac14\le d^\H(u,e)\le \frac12$.
Consider the set $A\subseteq B\times B$ consisting of pairs of
points which lie on a line segment joining a point $p\in
B_{\e^{1/9}}(e)$ and a point $q\in B_{\e^{1/9}}(u)$. Then $\mu\times
\mu(A)\gtrsim \e^{8/9}$, so that our assumption implies that
$$\frac{1}{\mu\times \mu(A)}\|d_\P-d_f\|_{L_1(A)}\le \e^{1/9}.$$
 By a Fubini argument it follows
that there exist $p\in B_{\e^{1/9}}(e)$ and $q\in B_{\e^{1/9}}(u)$
such that if we denote by $I=[p,q]$ the line segment joining $p$ and
$q$ then $$\|d_\P-d_f\|_{L_1(I\times I)}\lesssim \e^{1/9}.$$ Fix an
integer $n\approx \e^{-1/45}$. For $i\in \{0,\ldots,n\}$ let
$$J_i=\left[\frac{2i}{2n+1},\frac{2i+1}{2n+1}\right]\subseteq [0,1].$$
Then for every $(t_0,\ldots,t_n)\in J_0\times\cdots\times J_n$ the
additivity of $d_\P$ on the line segment $I$ implies that $$
d_P\left(v_{t_n},v_{t_0}\right)=\sum_{i=0}^{n-1}d_\P\left(v_{t_i},v_{t_{i+1}}\right).$$
Integrating this equality over $J_0\times\cdots\times J_n$ we get $$
\int_{J_n\times J_0} d_\P\left(v_s,v_t\right)dsdt=\sum_{i=0}^{n-1}
\int_{J_i\times J_{i+1}}d_\P\left(v_s,v_t\right)dsdt.$$ Since
$\|d_\P-d_f\|_{L_1(I\times I)}\lesssim \e^{1/9}$ it follows that
$$\sum_{i=0}^{n-1} \int_{J_i\times
J_{i+1}}d_f\left(v_s,v_t\right)dsdt\lesssim \int_{J_n\times J_0}
d_f\left(v_s,v_t\right)dsdt+n\e^{1/9}.$$ Assume that for all $i\in
\{0,\ldots,n-1\}$ and $(v_s,v_t)\in J_i\times J_{i+1}$ we have
$d_f(v_s,v_t)\ge \frac{\beta}{\sqrt{n}}$. Then using the fact that
$f$ is $1$-Lipschitz we arrive at the bound
$n\cdot\frac{1}{n^2}\frac{\beta}{\sqrt{n}}\lesssim
\frac{1}{n^2}+n\e^{1/9}$, and therefore $\beta\lesssim
n^{5/2}\e^{1/9}\lesssim \e^{1/18}$.

We proved above that there exists $i\in \{0,\ldots,n-1\}$ and
$(v_s,v_t)\in J_i\times J_{i+1}$ such that $$d_f(v_s,v_t)\lesssim
\frac{\e^{1/18}}{\sqrt{n}}.$$ Writing $v_s=(a_1,a_2,a_3)$ and
$v_t=(b_1,b_2,b_3)$ one checks that $|a_1-b_1|,|a_2-b_2|\lesssim
\e^{1/9}$ and $|a_3-b_3|\approx \frac{1}{\sqrt{n}}$. Therefore if we
set $w=(a_1,a_2,b_3)$ then $v_s$ and $w$ lie on the same coset of
the center and $d^\H(v_s,w)\approx \frac{1}{\sqrt{n}}\approx
\e^{1/90}$ while \begin{multline*}d_f(v_s,w)\lesssim d_f(v_s,v_t)+d_f(v_t,w)\lesssim
\frac{\e^{1/18}}{\sqrt{n}}+\e^{1/9}\\\lesssim \e^{1/18}d^\H(v_s,w),\end{multline*}
as required.

\section{Putting things together}\label{sec:finish}

Fix $\e>0$ and take $\delta=\e^K$ for a large enough $K>a$ that will
be determined presently, where $a$ is as in Theorem~\ref{thm:rate}.
Let $j$ and $y$ be as in Section~\ref{sec:choose monotone} for this
value of $\delta$, i.e., \eqref{eq:instead of lemma} is satisfied.
Thus $j\le \e^{-K}$. We now define $$M\coloneqq \left\{E\in
\cut\left(B\right):\ {\rm NM}_{B_{\delta^j}(y)}(E)\le \e^a\right\}.$$
Then by Markov's inequality applied to~\eqref{eq:instead of lemma}
we are ensured that $$\Sigma_f\left(\cut\left(B\right)\setminus
M\right)\lesssim \e^{K-a}\delta^j.$$ Define two semi-metrics on $B$
by $$d_1(p,q)\coloneqq \int_{ M}d_E(p,q)d\Sigma_f(E)$$ and
$$d_2(p,q)\coloneqq \int_{\cut\left(B\right)\setminus
M}d_E(p,q)d\Sigma_f(E)=d_f-d_1.$$ Then for all $p,q\in
B_{\delta^j}(y)$ we have $d_2(p,q)\lesssim \e^{K-a}\delta^j$.

By the definition of $M$, for all $E\in M$
Theorem~\ref{thm:stability} implies that there exists a half space
$\mathcal P_E$ for which
\begin{equation}\label{eq:symmetric difference}
\mu\left((E\cap B_{\e \delta^j}(x))\triangle \mathcal
P_E\right)\lesssim\e^{1/3}\left(\e\delta^j\right)^4.
\end{equation}
We shall now use the splitting of the cut measure from
Section~\ref{sec:split} with $r\coloneqq \e\delta^j$, $p=y$, and a
parameter $\theta>0$ which will be determined presently. Define two
semi-metrics on $B$ by $$d_3(u,v)\coloneqq \int_{M\cap
D_\theta}d_E(u,v)d\Sigma_f(E)$$ and $$\rho(u,v)\coloneqq \int_{M\cap
D_\theta}d_{\mathcal P_E}(u,v)d\Sigma_f(E)$$ (here $D_\theta\subseteq
\cut(B)$ is as in Section~\ref{sec:split}). Then
\begin{eqnarray}\label{pass to half space}
&&\!\!\!\!\!\!\!\!\!\!\!\!\!\!\!\!\!\nonumber\|d_3-\rho\|_{L_1(B_r(y)\times B_r(y))}\le \int_{M\cap
D_\theta}\\&&\!\!\!\!\!\!\!\!\!\!\left( \int_{B_r(y)\times
B_r(y)}\left||\1_{E}(u)-\1_{\mathcal
P_E}(u)|+|\1_{E}(v)-\1_{\mathcal
P_E}(v)|\right|\right)\nonumber d\Sigma_f(E)\\&\stackrel{\eqref{eq:symmetric
difference}}{\lesssim}& \Sigma_f(D_\theta)\e^{1/3}r^8\lesssim
\frac{\e^{1/3}r^8}{\theta},
\end{eqnarray}
where in the last inequality of~\eqref{pass to half space} we used
the bound $\Sigma_f(D_\theta)\lesssim \frac{1}{\theta}$ from
Section~\ref{sec:split}. Note that with $d_\theta$ as in
Section~\ref{sec:split} we have the point-wise inequality
$$|d_\theta-d_3|\le d_2\le \e^{K-a}\delta^j=\e^{K-a-1}r.$$

Now,
\begin{eqnarray}\label{eq:triangle}
&&\!\!\!\!\!\!\!\!\!\!\!\!\!\!\!\!\!\!\!\!\!\!\nonumber\frac{\|d_f-\rho\|_{L_1(B_r(y)\times
B_r(y))}}{\mu(B_r(y))^2}\\&\lesssim&
\frac{\|d_f-d_\theta\|_{L_1(B_r(y)\times
B_r(y))}}{r^8}+\frac{\|d_\theta-d_3\|_{L_1(B_r(y)\times
B_r(y))}}{r^8}\nonumber\\&\phantom{\le}&+\frac{\|d_3-\rho\|_{L_1(B_r(y)\times
B_r(y))}}{r^8}\nonumber\\&\stackrel{\eqref{eq:before sobolev}\wedge\eqref{pass
to half space}}{\lesssim}&\frac{r^{28/3}\theta^{1/3}+\e^{K-a-1}r\cdot
r^8+\e^{1/3}r^8\theta^{-1}}{r^8}\nonumber\\&=&r^{4/3}\theta^{1/3}+\frac{\e^{1/3}}{\theta}+\e^{K-a-1}r.
\end{eqnarray}
The optimal choice of $\theta$ in~\eqref{eq:triangle} is
$\theta\approx \frac{\e^{1/4}}{r}$. This yields the bound
\begin{equation*}\label{eq:L1 degeneracy}
\frac{\|d_f-\rho\|_{L_1(B_r(y)\times
B_r(y))}}{\mu(B_r(y))^2}\lesssim
r\left(\e^{1/12}+\e^{K-a-1}\right)\lesssim \e^{1/12}r,
\end{equation*}
provided that $K-a-1\ge \frac{1}{12}$. The result of
Section~\ref{sec:half} now implies that there exist $w,z\in B_r(y)$
which lie on the same coset of the center and
$$d^\H(w,z)\gtrsim
\e^{1/1080}r\quad  \mathrm{yet} \quad d_f(w,z)\lesssim \e^{1/216}d^\H(w,z).$$ Since
$j\le \frac{1}{\delta}$ and $\delta=\e^K$ we see that
$$d^\H(w,z)\gtrsim\e^{1/1080}\cdot \e\delta^j\ge \e^{2+K\e^{-K}}\ge
e^{-\e^{-2K}}$$ for $\e$ small enough. The proof of
Theorem~\ref{thm:rate} is complete.

\section{Concluding remarks}

We have presented here the complete details of the proof of Theorem~\ref{thm:main}, assuming only Theorem~\ref{thm:stability} on the stability of monotone sets, whose proof constitutes the bulk of~\cite{ckn}. The obvious significance of Theorem~\ref{thm:main} is that it shows that the correct asymptotic ``ballpark" of the integrality gap of the Sparsest Cut SDP is in the power of $\log n$ range. But, this result has other important features, the most notable of which is that it shows that the $L_1$ distortion of doubling, and hence also decomposable, $n$-point metric spaces can grow like $(\log n)^{\Omega(1)}$ (we refer to~\cite{LN06} for an explanation of the significance of this statement). Moreover, unlike the construction of~\cite{KV04} which was  tailored especially for this problem, the Heisenberg group is a classical and well understood object, which in a certain sense (which can be made precise), is the smallest possible $L_1$ non-embeddable metric space of negative type which posses certain symmetries (an invariant metric on a group that behaves well under dilations).

In addition to the above discussion, our proof contains several ideas and concepts which are of independent interest and might be useful elsewhere.
 Indeed, the monotonicity and metric differentiation approach to $L_1$-valued Lipschitz maps, as  announced (and sketched)
in Section 1.8 of~\cite{CK06}, was also  used in
 a much simpler form in~\cite{LR07}, in a combinatorial context and for a different purpose. Our proof is in a sense a ``hybrid" argument, which uses ideas from~\cite{CK06}, as well as the simplified proof in~\cite{ckmetmon}, with a crucial additional ingredient to estimate the scale. We prove a stability version of the classification of monotone sets in~\cite{ckmetmon}, but unlike~\cite{ckmetmon} we also need to work with perimeter bounds following~\cite{CK06} in order to deal with (using the isoperimetric inequality on $\H$ as in Section~\ref{sec:split}) the issue that the total mass of the cut measure does not have an a priori bound. In addition, the bound on the total perimeter is shown via the
kinematic formula
to lead to a bound on the total non-monotonicity, which in turn, leads
to the
scale estimate.


It is often the case in combinatorics and theoretical computer science that arguments which are most natural to discover and prove in the continuous domain need to be discretized. The ``vanilla approach" to such a discretization would be to follow the steps of the proof of the continuous/analytic theorem on the corresponding discrete object, while taking care to control various error terms that accumulate in the discrete setting, but previously did not appear in the continuous setting. An example of this type of argument can be found in~\cite{NS07}. Here we are forced to take a different path: we prove new continuous theorems, e.g. Theorems~\ref{thm:rate} and~\ref{thm:stability}, which yield ``rate" and ``stability" versions of the previously established qualitative theorems. Once such a task is carried out, passing to the required discrete version is often quite simple.

The need to prove stability versions of certain qualitative results is a recurring theme in geometric analysis and partial differentiation equations. As a recent example one can take the stability version of the isoperimetric theorem in $\R^n$ that was proved in~\cite{FMP08}. Another famous example of this type is the Sphere Theorem in Riemannian geometry (see~\cite{BS09} and the references therein).

In~\cite{ckn} we explain how our argument can be viewed as a general scheme for proving such results. The crucial point is to isolate a quantity which is
{\em coercive, monotone over scales, and admits an a-priori bound}. In our case this quantity is the total non-monotonicity. Coercivity refers to the fact that if this quantity vanishes then  a certain rigid (highly constrained) structure is enforced. Such a statement is called  a {\em rigidity} result, and in our setting it corresponds to the classification of monotone sets in~\cite{ckmetmon}. More generally (and often much harder to prove), the coercive quantity is required to have the following {\em almost rigidity} property: if it is less than $\epsilon^a$, for some $a\in (0,\infty)$,
then in a suitable sense,
the structure is $\epsilon$-close to the one which is forced by the $\epsilon=0$ case. In our setting this corresponds to Theorem~\ref{thm:stability}, and as is often the case, its proof is involved and requires insights that go beyond what is needed for the rigidity result. The monotonicity over scales refers to the decomposition~\eqref{eq:decompose w_j}, and the a priori bound~\eqref{eq:bound sum}, which is a consequence of the Lipschitz condition for $f$, implies, as in Section~\ref{sec:choose monotone}, the existence of a controlled scale at which the coercivity can be applied. We point out the general character of the
estimate for the scale thus obtained, which is the reason for the logarithmic behavior in Theorem~\ref{thm:rate}:
such an estimate for the scale
will appear {\it whenever} we are
dealing with a nonnegative quantity  which can
be written as a sum of nonnegative terms, one controlling each scale,
 such that there is a definite
bound on the sum of the
terms.  We call such a quantity {\it monotone over scales} to reflect the fact that the sum is nondecreasing
as we include more and more scales. As one example among very many, the framework that was sketched above can be applied in the context of~\cite{chco1}.


\bibliographystyle{IEEEtranS}

\bibliography{ellone1}

\end{document}